\documentclass[a4paper,11pt,aps,pra,amsmath,amssymb,amsfonts,superscriptaddress,notitlepage]{revtex4-1}

\usepackage{graphicx} 
\usepackage{hyperref}
\usepackage{lipsum}
\usepackage{color} 
\usepackage{lmodern} 
\usepackage{textcomp} 
\usepackage[T1]{fontenc} 

\DeclareMathOperator*{\Trace}{Tr}
\newcommand*{\ket}[1]{|{#1}\rangle}

\newcommand*{\mean}[1]{\mathinner{\langle{#1}\rangle}}

\newcommand*{\ketbra}[2]{\mathinner{|{#1}\rangle\langle{#2}|}}

\def\be{\begin{equation}}
\def\ee{\end{equation}}
\def\bes{\begin{equation*}}
\def\ees{\end{equation*}}

\newcommand{\gtwocross}{g^{(2)}_{fe}}
\newcommand{\gtwoselfa}{g^{(2)}_{ff}}
\newcommand{\gtwoselfb}{g^{(2)}_{ee}}
\newcommand{\Rabiangl}{\theta_r}
\newcommand{\Pin}{P_{\rm in}}
\newcommand{\Gammage}{\Gamma_{ge}}
\newcommand{\Gammaef}{\Gamma_{ef}}
\newcommand{\fac}{\xi}

\newcommand{\Omd}{\Omega}
\newcommand{\rhoML}{\rho_{\rm ML}}
\newcommand{\OmRabiEff}{\Omega_{\rm gf}}

\newcommand{\geph}{\hat e}
\newcommand{\efph}{\hat f}
\newcommand{\fh}{\hat f}
\newcommand{\eh}{\hat e}

\begin{document}
\title{Correlations and entanglement of microwave photons \\ emitted in a cascade decay}
\author{Simone Gasparinetti}
\email{gasimone@phys.ethz.ch}

\author{Marek~Pechal}

\author{Jean-Claude~Besse}

\author{Mintu Mondal}

\author{Christopher~Eichler}

\author{Andreas~Wallraff}
\affiliation{Department of Physics, ETH Zurich, CH-8093 Zurich, Switzerland}

\date{\today}

\maketitle

\textbf{
An excited emitter decays by radiating a photon into a quantized mode of the electromagnetic field, a process known as spontaneous emission \cite{Loudon2000}. If the emitter is driven to a higher excited state, it radiates multiple photons in a cascade decay.
Atomic \cite{Freedman1972,Aspect1981} and biexciton cascades \cite{Benson2000,Akopian2006,Stevenson2006,Dousse2010,Mueller2014} have been exploited as sources of polarization-entangled photon pairs.
Because the photons are emitted sequentially, their intensities are strongly correlated in time, as measured in a double-beam coincidence experiment \cite{Clauser1974,Loudon1980}.
Perhaps less intuitively, their \text{phases} can also be correlated, provided a single emitter is deterministically prepared into a superposition state, and the emitted radiation is detected in a phase-sensitive manner and with high efficiency. Here we have met these requirements by using a superconducting artificial atom, coherently driven to its second-excited state and decaying into a well-defined microwave mode. Our results highlight the coherent nature of cascade decay and demonstrate a novel protocol to generate entanglement between itinerant field modes.
}

\begin{figure}%
\includegraphics{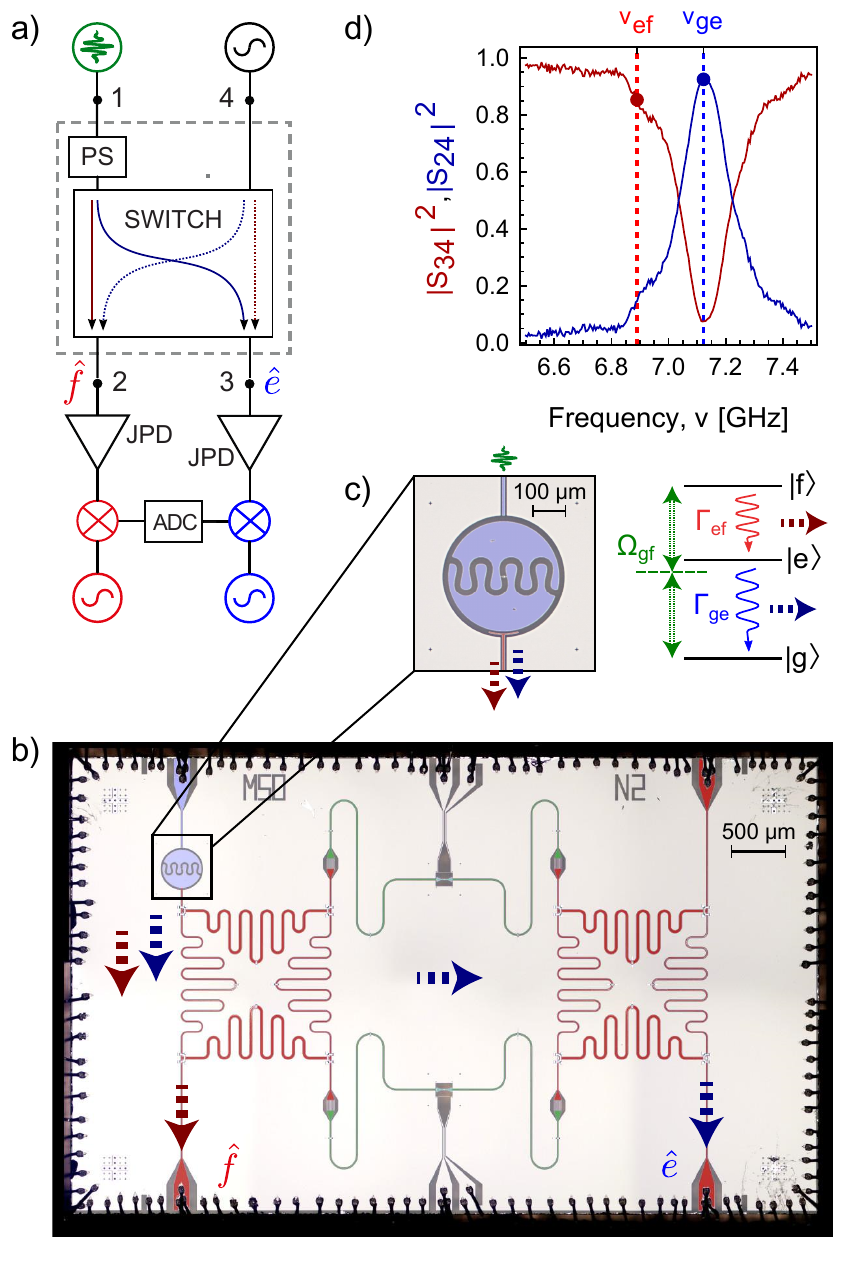}%
\caption{\textbf{Photon source and frequency-selective microwave switch.}
(a) Simplified scheme of the measurement setup. The photon source (PS) is coherently driven from input port 1. Input port 4 is used to characterize the scattering properties of the switch. The output modes $\geph$ and $\efph$ are amplified by Josephson parametric dimers (JPDs). The two quadratures of the amplified signals are measured by double heterodyne detection \cite{daSilva2010} and recorded by a digital-to-analog converter (DAC).
(b) False-color micrograph of the sample, illustrating the photon source (blue) and the key components of the switch, two $\pi/2$ hybrid couplers (red) and two tunable resonators (green).
(c) Micrograph of the photon source, consisting of a transmon \cite{Koch2007} asymmetrically coupled to input and output lines, and level scheme indicating the relevant transitions. The transmon is driven at the two-photon transition $\omega_{gf}/2$ with rate $\OmRabiEff$ and decays by emitting photons at frequencies $\omega_{ef}$ and $\omega_{ge}$ and rates $\Gammaef$ and $\Gammage$, respectively.
(d) Characterization of the switch via port 4: transmittance to modes $\hat e$ ($|S_{34}|^2$) and $\hat f$ ($|S_{24}|^2$) versus frequency $\nu$. The transition frequencies of the photon source, $\nu_{ge}$ and $\nu_{ef}$, are indicated by vertical dashed lines. 
}
\label{fig:setup}%
\end{figure}

We have realized a photon source (Fig.~\ref{fig:setup}) based on a transmon-type qubit \cite{Koch2007,Sathyamoorthy2016}, with two lowest transitions at 
$\omega_{ge}/2\pi=7.0975~\rm{GHz}$ and $\omega_{ef}/2\pi=6.8645~\rm{GHz}$ between the ground, $\ket{g}$, the first excited, $\ket{e}$, and the second excited state, $\ket{f}$, and a resulting anharmonicity $\alpha/2\pi=-233~\rm{MHz}$. The transmon is driven via a weakly coupled input port \cite{Peng2016d,Pechal2016} and decays into one of the two input ports of the single-pole, double-throw switch described in Ref.~\cite{Pechal2016}, with a measured rate $\Gammage/2\pi=(1.94\pm0.02)~\rm{MHz}$.
The switch has a tunable center frequency, which we set to $\omega_{ge}$, and a bandwidth of $150~\rm{MHz}$. Photons of frequency $\omega_{ge}$ ($\omega_{ef}$) fall within (out of) the switch bandwidth and are routed from the source to output mode $\hat e$ ($\hat f$) [Fig.~1(a)]. Each output mode is amplified using a nearly-quantum-limited Josephson parametric dimer (JPD) \cite{Eichler2014a} and its two quadratures are measured by heterodyne detection. We characterize the scattering properties of the switch by driving it via its second input port [Fig.~1(d)]. On this basis, we estimate that our routing scheme has an efficiency of $92\%$ ($86\%$) for photons of frequency $\omega_{ge}$ ($\omega_{ef}$).

We first investigate the cascade decay of our source under continuous-wave excitation. We coherently drive the two-photon transition between $\ket{g}$ and $\ket{f}$, with frequency $\omega_{gf}/2=(\omega_{ge}+\omega_{ef})/2$ [level scheme in Fig.~1(c)]. We vary the input power $P_{\rm in}$ and analyze the spectrum of the scattered radiation (Fig.~2). Contributions from coherent (Rayleigh) scattering at the drive frequency are discarded in our detection chain. At low powers, the spectrum shows a single peak at each of the transition frequencies $\omega_{ge}$ and $\omega_{ef}$. The photon flux into each mode, as given by the integrated power spectrum, first increases with power and then saturates as population is transferred into $\ket{f}$. 
At the same time, the two emission frequencies Stark-shift away from each other due to the driving of the two-photon transition 
\cite{Koshino2013}.
The measured data are in excellent agreement with a model based on a Lindblad-type master equation
and input-output theory (solid lines, see Methods). 
The Rabi frequency $\OmRabiEff$ describing coherent oscillations between $\ket{g}$ and $\ket{f}$ is proportional to the drive strength \textit{squared}, i.e., directly proportional to $P_{\rm in}$, because we are driving a two-photon transition.

Further insight into the emitted radiation is provided by its statistical properties \cite{Bozyigit2011,daSilva2010,Eichler2012,Lang2013}.
The power autocorrelation functions of the two modes, $g^{(2)}_{ee}(\tau)$ and $g^{(2)}_{ff}(\tau)$, show that the emitted radiation is antibunched \cite{SM}, a clear signature of single-photon emission \cite{Bozyigit2011,Hoi2012b,Peng2016d}.
Here, however, we focus on correlations \textit{between} the two photons, as expressed by the normalized power cross-correlation function $\gtwocross(\tau)=\mean{f^\dagger(0)e^\dagger(\tau) e(\tau) f(0)}/(\mean{e^\dagger e}\mean{f^\dagger f})$.
A positive (negative) time delay $\tau$ corresponds to a photon being emitted in mode $\geph$ after (before) a photon is emitted in mode $\efph$.
At the lowest drive powers [Fig.~2(a,b)], $\gtwocross(\tau)$ shows antibunching at negative time delays ($\gtwocross(\tau)\ll1$) and strong superbunching at positive ones ($\gtwocross(\tau)\gg1$), as expected for two time-correlated emissions occurring in a given sequence \cite{Clauser1974,Loudon1980}. This pattern changes at higher powers [Fig.~2(c,d)], specifically, when the pump rate becomes comparable to the decay rates ($\OmRabiEff/\Gammage \approx 1$). Due to fast repopulation of the $\ket{f}$ level, there is an increased probability for the reverse sequence to occur, which results in a superbunching peak appearing at negative time delays. 
More generally, $\gtwocross(\tau)$ develops oscillations at frequency $\OmRabiEff$, revealing the coherent nature of the population transfer between $\ket{g}$ and $\ket{f}$.
The observed properties are captured well by our model (solid lines), which has no free parameters and takes the $10~\rm{MHz}$ detection bandwidth of our setup into account (see Methods).

To characterize the radiation emitted in a single cascade, we use a preparation pulse, whose time envelope is a truncated Gaussian of variance $\sigma=5~\rm{ns}$ and controlled amplitude $A$.  The pulse drives coherent oscillations between $\ket{g}$ and $\ket{f}$, ideally preparing the transmon in the state
$\cos(\Rabiangl/2)\ket{g}+\sin(\Rabiangl/2)\ket{f}$, where $\Rabiangl$ is the preparation angle.
The optimal pulse duration is determined by a tradeoff between decay during state preparation and spectral overlap with the neighboring transitions at $\omega_{g}$ and $\omega_{ef}$, which are direct single-photon transitions.
We monitor the emitted radiation by performing ensemble averaging on the relevant moments of $e(t)$ and $f(t)$ \cite{daSilva2010,Eichler2012}.

We first consider the instantaneous powers $\mean{f^\dagger(t)f(t)}$ and $\mean{e^\dagger(t)e(t)}$, whose time dependence corresponds to the temporal shapes of the emitted photons [Fig.~\ref{fig:TimeRes}(a,d) and (b,e)]. The two shapes differ from each other: the power in the $\efph$ mode reaches its maximum at an earlier time and decays faster than in the $\geph$ mode. These features are explained by considering that (i) the decay rates into the two modes are different by a factor $\Gammaef/\Gammage\approx 2$, due to the dipole matrix elements of the transmon \cite{Koch2007}, and (ii) while the $\ket{f}$ state is directly populated by the excitation pulse, the $\ket{e}$ state is initially empty: the two photons decay sequentially.
The integrated powers, proportional to the photon numbers in each mode, oscillate as a function of the pulse amplitude squared [Fig.~\ref{fig:TimeRes}(g)]. These oscillations reflect those in the prepared $f$-state population, to which both photon numbers are proportional. We fit our data to a master equation simulation taking into account the time dependence of the preparation pulse as well as radiative decay during preparation [Fig.~\ref{fig:TimeRes}(g), solid lines], and use the fit to determine the preparation angle $\Rabiangl$ as a function of the pulse amplitude $A$.

We next consider the relative phase between the two photons. The amplitude-amplitude correlation $\mean{f(t)e(t)}$ is generally nonzero, indicating a well-defined relative phase [Fig.~\ref{fig:TimeRes}(c,f) and (h)]. The correlation is largest when the source is prepared in an equal state superposition ($\Rabiangl=\pi/2,3\pi/2$), and smallest when it is prepared in an energy eigenstate ($\Rabiangl=0,\pi$). By contrast, we have verified that the individual mode amplitudes $\mean{f(t)}$ and $\mean{e(t)}$ vanish identically, regardless of the pulse amplitude. 
We interpret the observed results in terms of a mapping of the transmon state into two itinerant photonic modes, $\hat e$ and $\hat f$. 
These modes are defined (and measured) by integrating the signals $e(t)$ and $f(t)$ over weighted time windows corresponding to the temporal shape of the emitted photons (temporal mode matching) \cite{Eichler2011} (see Methods).
In a Hilbert space comprising the transmon as well as the two modes, the cascade decay is described by the transformation
\be
\begin{split}
&\left[\cos(\Rabiangl/2)\ket{g}+\sin(\Rabiangl/2)\ket{f} \right] \otimes \ket{0_f} \otimes \ket{0_e}\\
 \to 
&\cos(\Rabiangl/2)\ket{g}\otimes\ket{0_f}\otimes\ket{0_e} +\sin(\Rabiangl/2)\ket{f}\otimes\ket{1_f}\otimes\ket{0_e} \\
 \to 
&\ket{g} \otimes \left[\cos(\Rabiangl/2)\ket{0_f}\otimes\ket{0_e} +\sin(\Rabiangl/2)\ket{1_f}\otimes\ket{1_e} \right] 
\end{split} \label{eq:entangle}
\ee 
where $\ket{0_{f,e}}$ and $\ket{1_{f,e}}$ indicate Fock states of the two modes with photon numbers zero and one. Equation \eqref{eq:entangle} stands for an entangling operation in which the superposition state created in the transmon is eventually shared by the two modes.

We fully characterize the two modes by performing joint tomography on them \cite{Eichler2011,Eichler2012} for the preparation angles $\Rabiangl=\pi/2$ and $\pi$ [Fig.~\ref{fig:hist4D}(a-b)]. The measured moments are referred to the two outputs of the switch using the calibration procedure described in the Methods.
The vanishing of first-order moments indicates that the radiation in each mode has no definite phase.
Photon-photon phase correlations manifest themselves in the amplitude correlation, $\mean{\fh\eh}$, which is substantial for a $\pi/2$ pulse and vanishes for a $\pi$ pulse.
The single-photon nature of the radiation is confirmed by the vanishing of
the fourth-order moments, $\mean{(\fh^\dagger)^2\fh^2}$ and $\mean{(\eh^\dagger)^2\eh^2}$.
Given the measured values for the moments and their respective standard deviations, we also determine the most likely density matrix $\rhoML$ describing the two modes at the output of the switch \cite{Welsch1999,Eichler2012,Lang2013}.
For a $\pi/2$ pulse [Fig.~\ref{fig:hist4D}(c)], we obtain a fidelity $F=\langle \psi | \rhoML | \psi \rangle = 91\%$ to the Bell state $\ket{\psi}=(\ket{00}+\ket{11})/\sqrt{2}$, and a negativity $\mathcal N(\rhoML) = -0.43$.

Our work demonstrates that in spite of its sequential nature, cascade decay must be regarded as a fully coherent process generating phase coherence between the emitted photons. Our excitation scheme can be extended to other architectures and generalized to multi-photon cascades and multiple modes. The ability to generate entanglement between spatially separated, itinerant radiation fields \cite{Eichler2011a,Flurin2012,Menzel2012,Lang2013,Peng2015b,Laehteenmaeki2016}, as demonstrated in our experiment, is essential to quantum information distribution protocols.


\clearpage

\begin{figure}%
\includegraphics{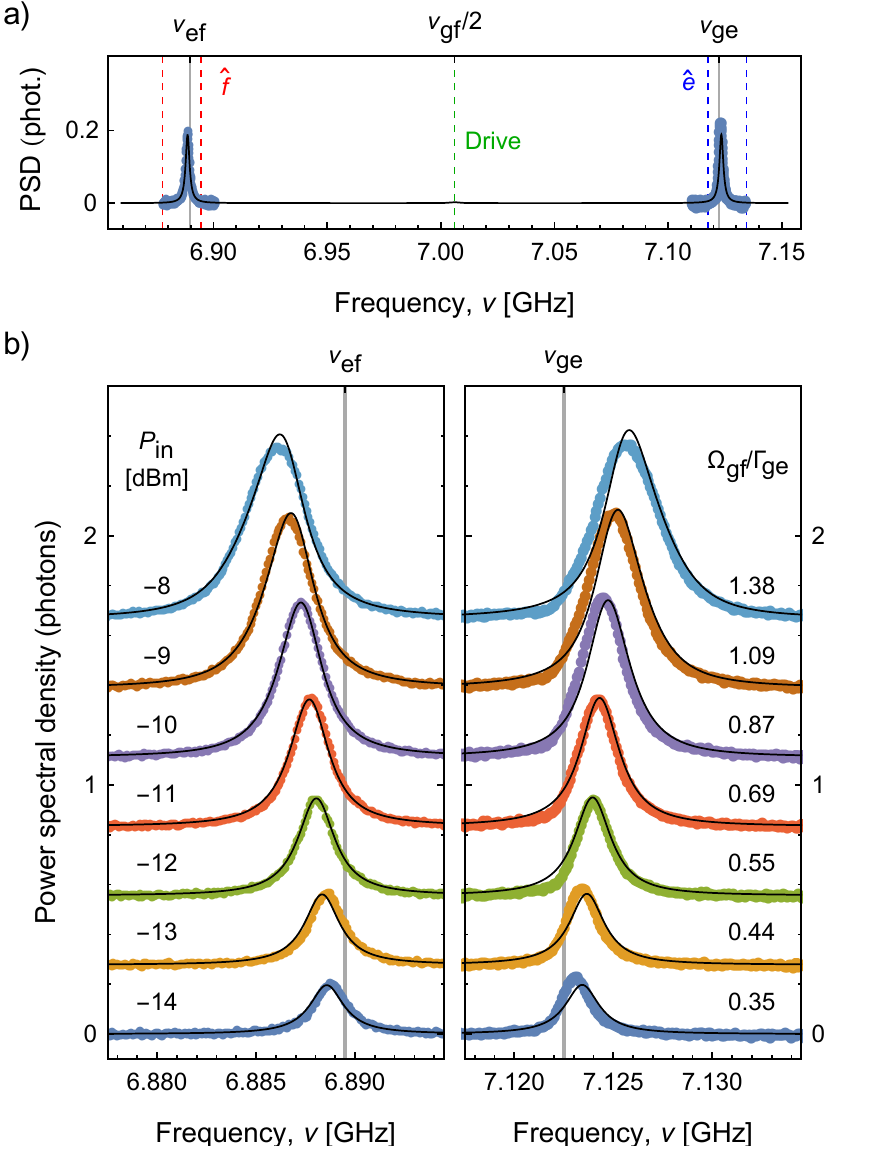}  
\caption{\textbf{Inelastic scattering under continuous excitation.}
(a) Power spectral density (PSD) of the radiation emitted by the source, driven at $\omega_{\rm gf}/2$ and low power: theory (solid line) and combined measured traces (colored dots) from the two modes. (b) Measured power spectral densities in modes $\efph$ (left) and $\geph$ (right) for varying input powers, $P_{\rm in}$. The traces are offset vertically for clarity and globally fitted to our model (solid lines, see Methods). For each input power $\Pin$, we note the corresponding $\ket{g}\leftrightarrow \ket{f}$ drive rate $\OmRabiEff$, relative to the decay rate $\Gammage$.
}%
\label{fig:psds}%
\end{figure}

\begin{figure}%
\includegraphics{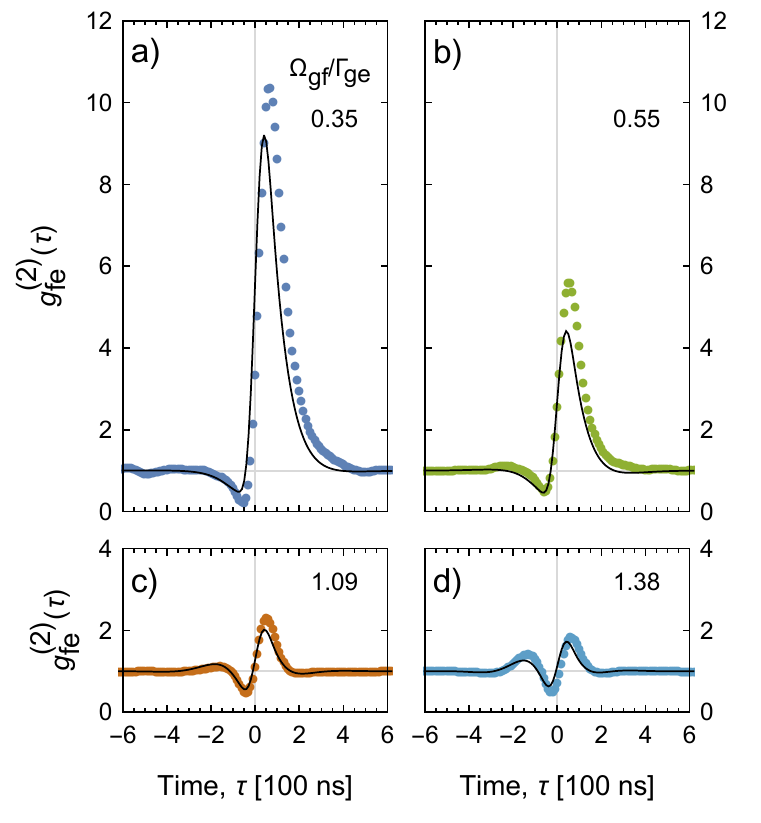}%
\caption{
\textbf{Photon-photon correlations.}
Time-resolved power cross-correlation $\gtwocross(\tau)$ between modes $\hat f$ and $\hat e$, taken at the  indicated, normalized drive rates $\OmRabiEff/\Gammage$ (dots). The solid lines are calculated based on the parameters of Fig.~\ref{fig:psds}, taking into account the $10~\rm{MHz}$ detection bandwidth of the used experimental setup (see Methods) and with no fit parameters.
}%
\label{fig:g2cross}%
\end{figure}

\begin{figure}%
\includegraphics[width=\linewidth]{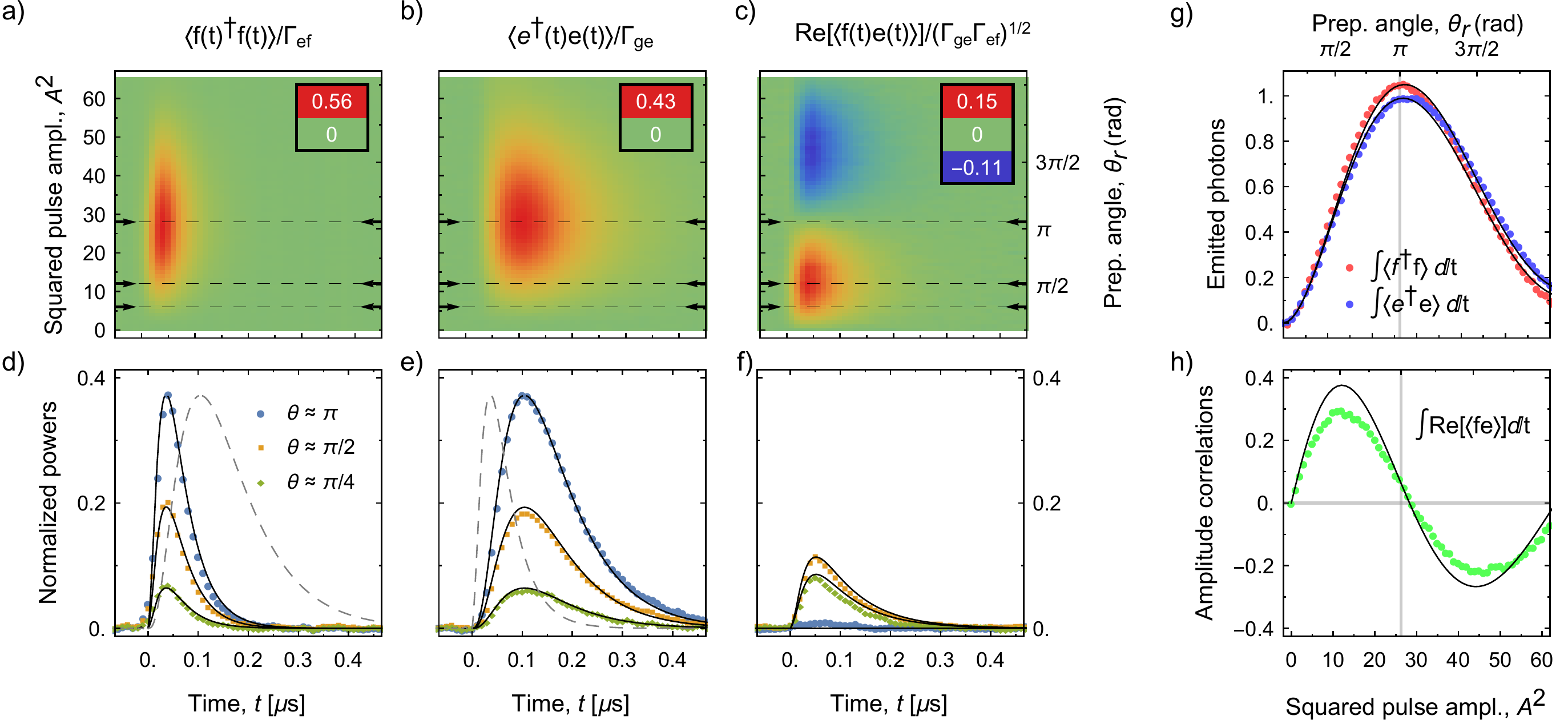}%
\caption{\textbf{Pulsed excitation: photon shapes and phase correlations.}
(a-c)
Density plots of the normalized instantaneous power in each mode, $\mean{f^\dagger(t) f(t)}/\Gammaef$ (a) and $\mean{e^\dagger(t) e(t)}/\Gammage$ (b), and of the real part of the normalized amplitude cross-correlation $\mean{f(t) e(t)}/(\Gammage \Gammaef)^{1/2}$ (c), versus time, $t$ (horizontal axis), and squared amplitude of the excitation pulse, $A^2$ (left vertical axis). The right vertical axis shows the preparation angle $\Rabiangl$, calibrated using the data of panel (g).
(d-f) Data (colored dots) of (a-c) extracted along the dashed lines. Solid lines are calculations based on the measured parameters of the transmon, the preparation angle $\Rabiangl$, and the measured bandwidth of each amplification chain (see Methods). In panels (d,e) we have also plotted a reference trace from the other panel (dashed lines). (g) Integrated powers, $\int \mean{f^\dagger f} dt$ and $\int \mean{e^\dagger e} dt$, versus $A^2$. The solid lines are a fit to the theory model, serving as a calibration for the preparation angle $\Rabiangl$. (h) Integrated amplitude cross correlation, $\int\mean{fe} dt$, versus $A^2$: data (dots) and corresponding theory (solid line) for the same parameters as in (g).}%
\label{fig:TimeRes}%
\end{figure}

\begin{figure}%
\includegraphics{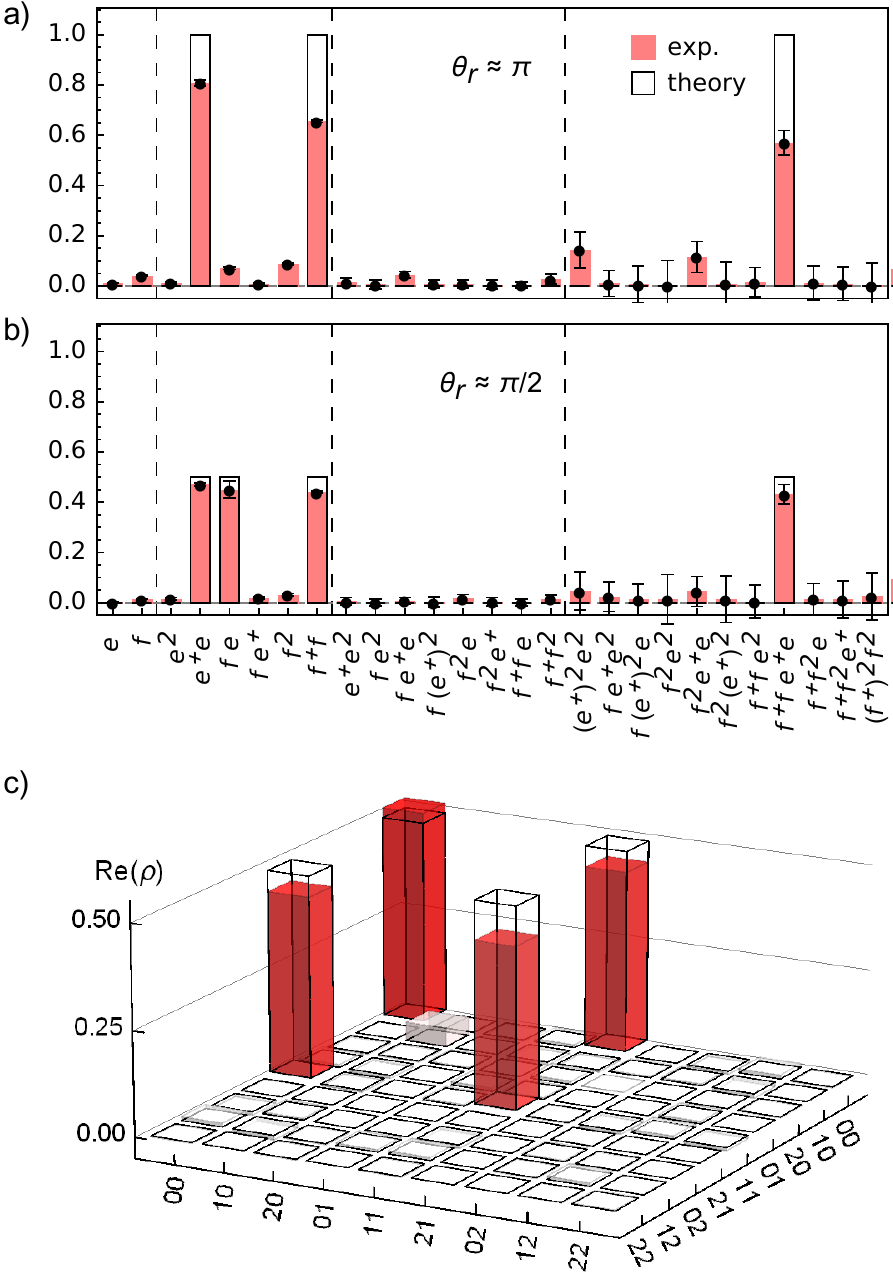}
\caption{\textbf{Joint quantum state tomography of the modes $\hat e$ and $\hat f$.} (a-b) Reconstructed moments of the temporally matched modes $\hat f$ and $\hat e$ (absolute values) up to fourth order for the preparation angles (a) $\Rabiangl\approx\pi$ and (b) $\Rabiangl\approx\pi/2$ (color bars).
The moments are referenced to the output of the switch (see Methods).
The expected moments for the states $\ket{11}$ and $(\ket{00}+\ket{11})/\sqrt{2}$ are shown as wireframes in (a) and (b), respectively. (c) Real part of the most likely density matrix corresponding to the measured moments in (b), displaying a $91\%$ fidelity to the state $(\ket{00}+\ket{11})/\sqrt{2}$ (wireframes) and a negativity of $-0.43$. All entries of the imaginary part (not shown) have absolute value below $0.01$. 
}%
\label{fig:hist4D}%
\end{figure}


\clearpage
\section*{Methods}

\textbf{Experimental details.} Measurements are carried out in a dilution refrigerator at 35~mK. The measured device is the same as in Ref.~\cite{Pechal2016}. The frequency of the transmon and the working point of the switch are set by magnetic-flux tuning, for which we use two on-chip flux lines and a superconducting coil mounted underneath the chip. The transition frequencies and linewidth of the transmon are measured by low-power spectroscopy. In Fig.~1(d), the measurement of the scattering parameters is referenced to a detuned configuration of the switch in order to compensate for frequency-dependent attenuation and gain in the measurement lines. In Fig.~2, power spectral densities are measured as Fourier transforms of the autocorrelation functions of the two signals. In Fig.~3, power correlations are measured by double heterodyne detection followed by real-time signal processing prior to ensemble averaging \cite{Bozyigit2011,daSilva2010,Lang2013}.
The noise added by each detection chain is characterized by measuring its statistical properties with the drive tone turned off, and subtracted from the signal using the methods described in Refs.~\cite{daSilva2010,Eichler2012}. Our detection bandwidths are set by the bandwidth of the parametric amplifiers and by the digital filter applied to the signals. In the measurements of Fig.~2, the detection bandwidth is $20~\rm{MHz}$ for both modes. In those of Fig.~3, it is $10~\rm{MHz}$ for both modes (limited by digital filtering). In those of Fig.~4, it is $22~\rm{MHz}$ for mode $\hat f$ and $11~\rm{MHz}$ for mode $\hat e$ (limited by the parametric amplifiers). The measurements of Fig.~5 are taken by integrating the two output signals with an appropriate mode-matching filter. Due to technical reasons, we are restricted to using the same filter for both channels; we therefore define a filter whose shape is intermediate between that of the two modes [as determined from Fig.~4(d,e)], and optimize the delay between the integration windows (see also the Supplementary Material \cite{SM}).

\textbf{Theoretical modeling.} Our model is based on a Lindblad-type master equation for a driven, ladder-type, three-level system coupled to a semi-infinite transmission line. The ratio between the dipole moments of the $ef$-- to the $ge$--transition is taken to be $\fac=\sqrt{2}$ \cite{Koch2007}. The output fields and their correlations are calculated using input-output theory \cite{Walls2008} and the quantum regression theorem \cite{Carmichael2002}. More details can be found in the Supplementary Material \cite{SM}.

\textbf{Data analysis.}
In Fig.~2, the measured spectra are globally fit to our model, the only fit parameter being the total attenuation of the drive line (input 1 in Fig.~1).

In Fig.~3, we incorporate the finite detection bandwidth by convolving the calculated $\gtwocross(\tau)$ with the squared kernel of the used digital filter twice. This is an approximation that holds in the limit of low signal-to-noise ratio relevant here; for more details, see Ref.~\cite{Lang2014}.

In Fig.~4(g), the solid lines are obtained by numerically solving the master equation with a time-dependent drive whose envelope matches our preparation pulse, and using the solution to calculate the integrated powers in the two modes as a function of the pulse amplitude. We fit the theory to the data and extract the conversion factor between pulse amplitude and drive strength, which we use to estimate the preparation angle $\Rabiangl$. To model the data of Fig.~4(d-f), we assume (for simplicity) instantaneous state preparation of the transmon at time $t=0$. We first calculate the two-time correlation functions $\mean{x^\dagger(t) y(t+\tau)}$, with $\{x,y\}=\{e,f\}$, and then convolve these functions with Lorentzian kernels corresponding to the measured detection bandwidths.

The moments in Fig.~5(a,b) are reconstructed as explained in Ref.~\cite{Eichler2012} and referred to the outputs of the switch by applying global scaling factors to them.
According to the master equation simulation described in the Supplementary Material \cite{SM}, a preparation angle $\theta=\pi/2$ results in an average photon number of $0.50$ in both modes.
After taking into account inefficient signal routing due to photon leakage into input port 1 ($2\%$, see Ref.~\cite{Pechal2016}) and imperfect frequency selectivity of the switch [see Fig.~1(d)], we estimate photon numbers $\mean{f^\dagger f}=0.42$ and $\mean{e^\dagger e}=0.45$ at the output of the switch, which we have used to scale the moments shown in Fig.~5. This scaling corresponds to total detection efficiencies $\eta_f=0.07$ for mode $\hat f$ and $\eta_e=0.08$ for mode $\hat e$, which we believe to be limited by cable losses, noise added by the amplifiers, and imperfect temporal mode matching (see also Supplementary Information \cite{SM}).
The error bars in Fig.~5(a,b) are determined as the standard deviation of 20 repeated measurements, each of which was averaged 64M times.
In Fig.~5(c), we estimate the most likely density matrix using the algorithm described in Ref.~\cite{Eichler2012} and applied, for instance, in Ref.~\cite{Lang2013}. We restrict the Hilbert space to up to two photons in each mode.

In the Supplementary Figure S1, the data are corrected for weak thermal background radiation during the reference (``off'') measurements \cite{Lang2013,Eichler2015}. The correction corresponds to a thermal population $n_{\rm th}\approx0.01$ in mode $\efph$ and $n_{\rm th}\approx0.005$ in mode $\geph$.

\bigskip

\clearpage
%

\clearpage
\section*{Supplementary Information}

\renewcommand\thefigure{S\arabic{figure}}    
\setcounter{figure}{0}   
\renewcommand\theequation{S\arabic{equation}}    
\setcounter{equation}{0}   

\section{Supplementary data}


\subsection{Photon antibunching}

In Fig.~S1 we present measurements of the normalized power autocorrelation functions for the modes $\hat e$ and $\hat f$, $g^{(2)}_{aa}(\tau)=\mean{\hat a^\dagger(0) \hat a^\dagger(\tau)\hat a(\tau)  \hat a(0)}/(\mean{\hat a^\dagger \hat a})^2$ with $a=\{e,f\}$, taken under the same experimental conditions as the measurements of Fig.~2(d). Both functions exhibit clear antibunching, in a similar way as previously observed for single-photon sources \cite{Bozyigit2011}. Differently from single-photon emission by a two-level system, here we observe a flattening of the first-order derivative at small time delays. This effect, captured by our theory, is due to small oscillations in the autocorrelation functions, smoothed out by our digital filter.

\begin{figure}[b]%
\includegraphics[width=10cm]{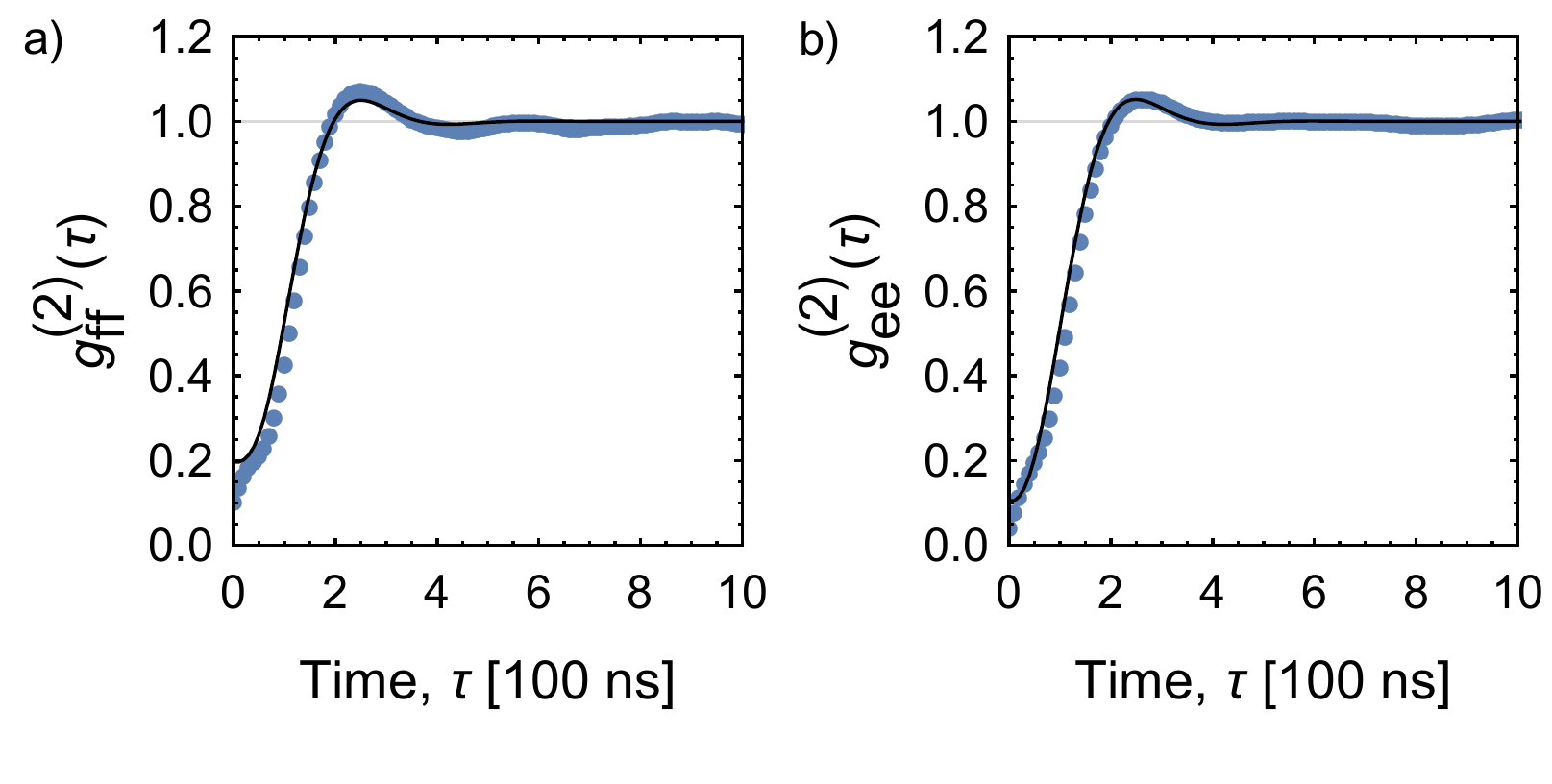}%
\caption{\textbf{Photon antibunching.}
Time-resolved power autocorrelation functions (a) $\gtwoselfa(\tau)$ and (b) $\gtwoselfb(\tau)$ of the two modes, measured at the Rabi rate $\OmRabiEff/2\pi=1.38$.
The solid lines are calculated using the model discussed in the Methods.
}%
\label{fig:g2self}%
\end{figure}

\section{Theoretical model}

\subsection{Master equation}

Our model is based on a master equation for a driven, ladder-type, three-level system coupled to a transmission line. The system Hamiltonian in a frame rotating at the drive frequency $\omega_d$ and after the rotating-wave approximation is
\be
H=\left(
\begin{array}{ccc}
 0 & \frac{\Omd }{2} & 0 \\
 \frac{\Omd }{2} & \frac{1}{2} (-\alpha -2 \delta ) & \frac{1}{2} \fac \Omd  \\
 0 & \frac{1}{2} \fac \Omd  & -2 \delta  \\
\end{array}
\right)
\label{eq:Ham}
\ee
where $\delta=\omega_d-\omega_{gf}/2$ is the detuning of the drive from the bare two-photon transition frequency, $\Omd$ is the drive strength, and $\fac$ is the relative strength between the two dipole transitions. For the transmon, $\fac\approx\sqrt{2}$ \cite{Koch2007}.
We write the master equation for the density matrix $\hat \rho(t)$ of the transmon as $\partial_t \hat\rho(t) = \mathcal L \hat\rho(t)$. The Liouvillian is given by
\be
\mathcal L \hat\rho(t) = -\frac{i}{\hbar}[\hat H,\hat\rho(t)]+ \Gammage \mathcal D[\hat\sigma_-^{T},\hat\rho(t)] \ , \label{eq:liou}
\ee
where $\Gammage$ is the decay rate of $\ket{e}$ into the transmission line. We have also introduced a total annihilation operator $\hat\sigma_-^{T}=\hat\sigma_-^{\rm ge}+\fac\hat\sigma_-^{\rm ef}$ comprising the individual annihilation operators for the two transitions, $\hat\sigma_-^{\rm ef}=\ketbra{e}{f}$, and $\hat\sigma_-^{\rm ge}=\ketbra{g}{e}$, and the dissipator is expressed in the standard form
\be
\mathcal D[A,B] = A BA^\dagger - \frac12\left(A^\dagger A B- B A^\dagger A\right) \ .
\ee

\subsection{Analytic expression for the Rabi frequency}

We consider the unitary evolution given by the Hamiltonian \eqref{eq:Ham} in the limit $\Omega \ll \alpha$ and with $\fac=\sqrt{2}$. The evolution operator can be expressed analytically in terms of roots of a cubic equation. For a given $\Omega$, we seek the optimal detuning that maximizes the visibility of oscillations between states $\ket{g}$ and $\ket{f}$. We find that unit visibility is attained for $\delta_{\rm opt}=\Omega^2/(4\alpha)$. This frequency shift from the bare two-photon resonance is an ac-Stark shift induced by the driving field. The corresponding Rabi frequency is given by $\OmRabiEff=\sqrt{2} \Omega^2/\alpha$, up to second order in the small parameter $\epsilon=\Omega/\alpha$. Numerical simulations confirm that this approximation is accurate in the power range expored in our experiment, for which $\epsilon<0.09$.

\subsection{Input-output theory}

Because the detection bandwidth of modes $\hat e$ and $\hat f$ is much smaller than the frequency difference between the two transitions (anharmonicity), we assume that the transmon decays from $\ket{f}$ into $\ket{e}$ by emitting a photon into the mode $\hat f$ (with rate $\Gammage$), and from $\ket{e}$ to $\ket{g}$ into mode $\hat e$ (with rate $\Gammaef=\fac^2 \Gammage$). The mode operators are thus given by
\be
\begin{split}
\hat f &= \sqrt{\Gammaef} \ \hat \sigma_-^{\rm ef} \ , \\
\hat e &= \sqrt{\Gammage} \ \hat \sigma_-^{\rm ge} \ .
\end{split}\label{eq:modes}
\ee

\subsection{Continuous excitation}

To model the response under continuous excitation, we first find the steady-state $\hat \rho_{\rm st}$ that satisifes $\mathcal L \hat\rho_{\rm st}=0$. We calculate the power spectral density of the emitted radiation for each mode $\hat a = \{ \hat e, \hat f\}$ (in units of photon flux per unit frequency, or simply photons) as the Fourier transform of the corresponding autocorrelation function:
\be
S_{a}(\omega)= \frac1{2\pi} \int_{-\infty}^\infty d\tau e^{i \omega \tau} \Trace[\hat a^\dagger e^{\mathcal L \tau} (\hat a \hat\rho_{\rm st})] \ .
\ee

The normalized power cross correlation between the modes is given by \cite{Carmichael2002}
\be
\gtwocross(\tau) = \Trace [\hat\sigma_+^{\rm ge} \hat\sigma_-^{\rm ge} e^{\mathcal L \tau}  (\hat\sigma_-^{\rm ef} \hat\rho_{\rm st} \hat\sigma_+^{\rm ef})] \ .
\ee

\subsection{Pulsed excitation}

The full dynamics of the system under pulsed excitation is obtained from a numerical solution of the master equation generated by \eqref{eq:liou}, with a time-dependent drive strength $\Omega(t)$. To obtain analytical expressions for the time dependence of the field moments, we take the limit of instantaneous state preparation and assume that at time $t=0$ the transmon is prepared in the state $\cos(\theta_r/2) \ket{g}+\sin(\theta_r/2)\ket{f}$.
Using Eqs.~\eqref{eq:modes}, we find that the amplitudes of each mode identically vanish: $\mean{f(t)}=0$ and $\mean{e(t)}=0$ at all times. The instantaneous powers read
\begin{align}
\mean{f^\dagger(t)f(t)} &=\Gammaef \sin ^2(\Rabiangl/2) e^{-\Gammaef t} \ , \\
\mean{e^\dagger(t)e(t)} &= \Gammage \sin ^2(\Rabiangl /2) \frac{e^{-\Gammage  t }-e^{- \Gammaef t}}{1-\Gammage/\Gammaef} \ .
\end{align}
In both modes, the power is proportional to the initial $f$-state population $P_f=\sin^2(\Rabiangl/2)$. In mode $\efph$, the power is maximum at the time of preparation and decays exponentially as the state $\ket{f}$ decays into $\ket{e}$ with rate $\Gammaef$. The nonmonotonic time dependence of the power in mode $\geph$ stems from two competing effects, as the emitting state $\ket{e}$, initially empty, is both populated by the decay from $\ket{f}$ and depleted by the decay into $\ket{g}$. As a result, the maximum power is reached at the delayed time
$\bar{t}_{ge}=2\ln(\Gammaef/\Gammage)/(\Gammaef-\Gammage)$
. Furthermore, the decay in mode $\geph$ is slower by a factor 
$\Gammaef/\Gammage$. 
To obtain the number of photons radiated into each mode, we integrate the two expressions and find
\be
\int_0^\infty\mean{f^\dagger(t)f(t)}dt = \int_0^\infty\mean{e^\dagger(t)e(t)}dt = \sin^2(\Rabiangl/2) \ .
\ee
The amplitude correlation between the two channels reads
\be
\mean{f(t)e(t)}= (\Gammaef/2) \sin \Rabiangl e^{- \Gammaef t/2} \ .
\ee
We note that this correlation is maximum at $t=0$, despite the fact that the $\ket{e}$ state is empty at that time. The integrated correlation reads
\be
\int_0^\infty\mean{f(t)e(t)}dt = \sin \Rabiangl \ .
\ee
The dependence of the integrated quantities on the Rabi angle $\Rabiangl$ is consistent with the coherent mapping of the state of the three-level system into two itinerant bosonic modes, according to Eq.~(1) in the main text.

\clearpage
\section{Experimental details}

\subsection{Loss in detection efficiency due to imperfect temporal mode matching}

For each time-dependent mode $\hat a(t)$ (with $a=\{e,f\}$), we define the corresponding temporally matched, time-independent mode $\hat a$ as $\hat a = \int dt \hat a(t) k(t)$,
where the mode-matching function $k(t)$ satisfies the normalization condition $\int dt |k(t)|^2=1$.

In our experiment, temporal mode matching is achieved by convolving the digitized signals with a finite-impulse response (FIR) filter and recording the result of the convolution at the time that maximizes the signal. Recalling the definition of convolution, one sees that the kernel of the filter corresponds to a time-reversed sampling of $k(t)$ at the digitization rate of the acquisition system. The optimal filter kernels for the two modes can be estimated from the measured shapes of the emitted photons [Fig.~4 and S2]. Due to the way our digital acquisition card is configured, here we are restriced to using the same filter for both modes. Imperfect mode matching results in a decrease in the overall detection efficiencies for the two modes. It does not, however, alter the statistical properties of the modes \cite{Eichler2012b}. We estimate the detection efficiency $\eta^{\rm mm}$ due to mode matching by convolving the filter with the square root of the measured photon shape. For the optimal filters, $\eta^{\rm mm}=1$. For the used filter, $\eta^{\rm mm}=0.95$ for mode $\hat f$ and $\eta^{\rm mm}=0.70$ for mode $\hat e$. 

\begin{figure}[b]%
\includegraphics[width=8.6cm]{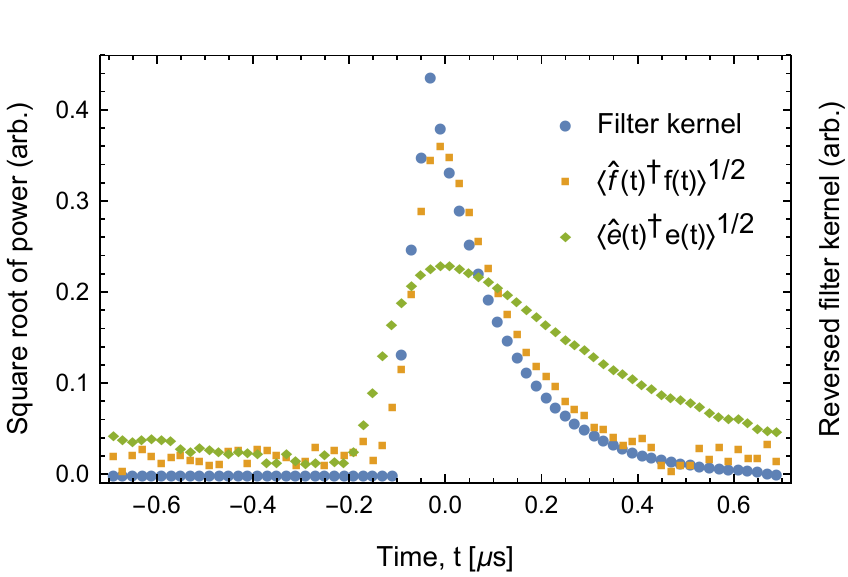}%
\caption{\textbf{Imperfect temporal mode matching.}
Time-shifted square root of the measured signal powers in the two modes, and time-reversed kernel of the filter used in the experiment.
}%
\label{fig:TimeResFilters}%
\end{figure}

\end{document}